\documentclass[journal=apchd5,manuscript=article]{achemso}

\usepackage{chemformula} 
\usepackage[T1]{fontenc} 
\usepackage{amsmath,amsfonts}

\author{Sreeraj Rajan Warrier}
\affiliation{Department of Physics, Mahindra University, Hyderabad-500043, Telangana, India}
\author{Jayasri Dontabhaktuni}
\affiliation{Department of Physics, Mahindra University, Hyderabad-500043, Telangana, India}
\email{jayasri.d@mahindrauniversity.edu.in}

\title{Inverse Design using Physics-Informed Quantum GANs for Tailored Absorption in Dielectric Metasurfaces}


\begin{document}

\begin{tocentry}

\includegraphics[width=8.2cm, height=4.5cm]{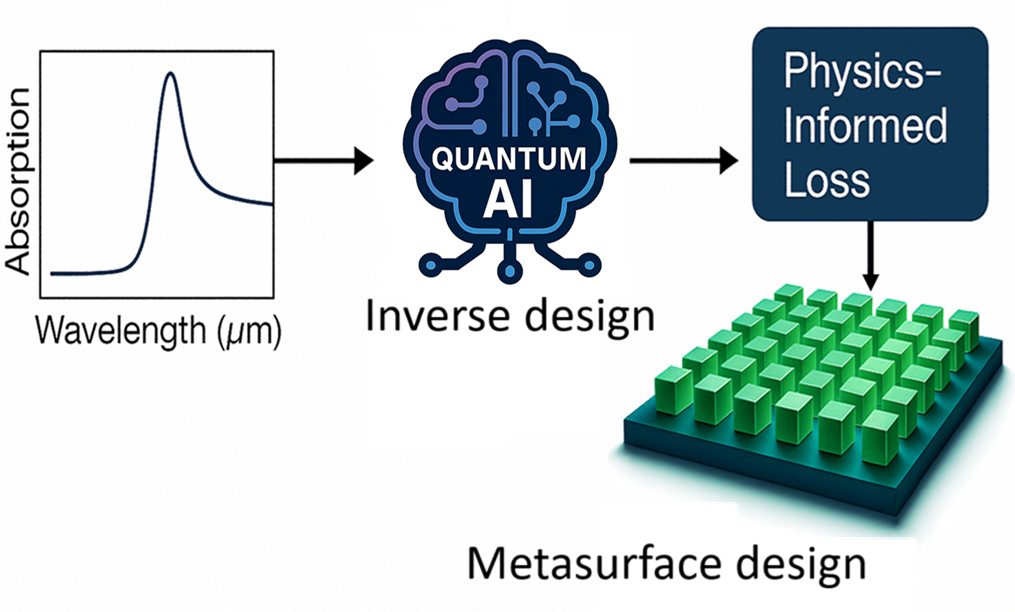}


\end{tocentry}

\begin{abstract}
High Q-factor narrow-band absorption exhibits high spectral selectivity enabling high-sensitive photodetectors, sensors and thermal emitters. All-dielectric metasurfaces are widely regarded as excellent candidates for giving rise to such narrow-band absorption. However, designing metasurfaces with specific functionalities remains a challenging task both experimentally and computationally, which is why inverse design methods are increasingly being explored. Inverse design process is highly complex due to its non-unique solutions and the higher dimensionality of the design space, making it challenging to precisely control the resonance wavelength, linewidth, and absorption intensity. In this paper, we present a novel hybrid methodology that integrates generative adversarial networks (GANs) (both classical and quantum) with physics-informed neural networks (PINNs) for the inverse design of narrow-band absorbing metasurfaces. By introducing a Fano-shaped absorption spectrum equation into the PINN loss function, we enforce physical constraints on the resonance behavior, ensuring outputs that are both spectrally accurate and physically consistent. The study presents a comparison between a conventional GAN + PINN framework and a PINN augmented by a hybrid quantum-classical GAN (QGAN). The findings indicate that the integrated PINN + QGAN model achieves faster convergence, requires 99.5\% fewer training samples, and yields an order of magnitude lower MSE compared to conventional GANs. Remarkably, even though the training dataset only contains metasurfaces with Q-factors on the order of $10^3$, the model is able to generate highly asymmetric metasurface structures with Q-factors exceeding $10^5$. This study presents a novel framework that integrates quantum machine learning with physics-based modeling, providing a promising method for quantum-enhanced inverse design in nanophotonic systems.
\end{abstract}

\section{Introduction}

Metasurfaces — planar arrangements of subwavelength nanostructures offer unprecedented control over the spatial, spectral, and polarization characteristics of electromagnetic waves. Among their many functionalities, achieving narrow-band absorption is of critical importance for a variety of applications that require high spectral selectivity. For instance, in thermal emitters, narrow-band absorbers enable wavelength-specific emission following Kirchhoff’s law, which is vital for thermophotovoltaics and infrared camouflage~\cite{Liu2011}. In sensing, sharp absorption resonances enhance the interaction of light with molecular vibrational modes, thereby improving sensitivity and specificity in spectroscopic detection~\cite{li2021metasurface}. Narrow-band absorbers also play a key role in photodetection, where they improve the signal-to-noise ratio by suppressing unwanted spectral components~\cite{Liang2022}. Furthermore, their ability to support enhanced refractive index sensing makes them suitable for applications in optical filters, modulators, and biosensing platforms~\cite{cheng2023terahertz}. Thus, engineering narrow-band absorption in metasurfaces is fundamental to advancing next-generation nanophotonic devices.
A key functionality in many of these applications is the ability to tailor narrow-band absorption profiles with precise control over resonance frequency, linewidth, and peak amplitude~\cite{Chen2019}. 

Conventional forward simulations typically rely on computationally intensive methods such as the finite-difference time-domain (FDTD) or finite element method (FEM) to solve Maxwell's equations. While these approaches provide accurate results, they often struggle with scalability and efficiency when exploring large design spaces or optimizing complex metasurface structures. Of late inverse design methods are being increasingly explored for tasks where one seeks to obtain a desired electromagnetic response from a metasurface geometry. Methods based on topology optimization, Bayesian optimization, machine and deep learning methods are investigated for performing the inverse design of metamaterials. More specifically algorithms such as generative adversarial networks (GANs), are being utilized to establish mappings between geometric configurations and the desired optical responses \cite{Yeung2021global}. Classical GANs are capable of generating a variety of candidate designs; however, they generally require substantial training data and do not inherently adhere to physical laws, which can lead to outputs that are physically inconsistent and may suffer from mode collapse~\cite{Yeung2021global}. Furthermore, designing metasurfaces to meet stringent specifications presents inherent complexities, stemming from the high-dimensional, nonlinear, and non-unique characteristics of the inverse design problem. 

Physics-informed neural networks (PINNs) have been developed to enhance generalization and physical reliability by integrating physical models directly into the training process through the use of customized loss functions~\cite{raissi2019physics, chen2020physics}. PINNs in metasurface design employ analytical frameworks, such as coupled mode theory, to direct the learning process towards designs that adhere to electromagnetic limitations~\cite{tanriover2020physics, Yucheng2024}. This enables the construction of more physically feasible and effective structures. Research indicates that incorporating physical information into neural networks (NNs) helps alleviate their training demands and address the constraints of conventional NNs, particularly when explicit governing equations exist for the issues \cite{Yucheng2024, chen2020physics}. Typically, inverse design of metasurfaces include the physical information - by altering the model architecture, including explicit physical equations into loss functions, or by adjusting model parameters. However, majority of these studies depend on the possession of a precise spectrum from the beginning, which is sometimes unfeasible in real applications.

The emergence of quantum machine learning concurrently introduces new opportunities for generative modeling. Quantum computing offers a distinctive approach for the inverse design of metasurfaces, particularly via the utilization of Quantum Generative Adversarial Networks (QGANs).  QGANs exhibit considerable potential to improve the design process by enabling efficient navigation and optimization of parameter spaces~\cite{preskill2018quantum, lloyd2018quantum}.  Integrating quantum computing with generative adversarial networks (GANs) enables faster and more efficient convergence toward optimal solutions, thereby advancing the development of precise and innovative metasurface designs across various technological fields~\cite{dallaire2018quantum}. QGANs utilize variational quantum circuits (VQCs) as part of a hybrid quantum-classical optimization framework. The ability to represent complex distributions using fewer parameters can significantly enhance expressiveness. These characteristics are particularly advantageous for inverse design problems, where data is limited or the solution space is highly complex and interconnected. Recent studies indicate that quantum generative algorithms could provide an exponential advantage compared to classical algorithms, resulting in increased interest in the theory and experimentation related to quantum GANs~\cite{Huang2021, Tsang2023, Stein2021, Romero2021, chang2024latent}.

In this study, we introduce a novel framework that combines physics-informed learning with both classical and quantum generative adversarial models for the inverse design of narrow-band absorbing metasurfaces. While many previous works have used PINNs to perform inverse design of metasurfaces, our approach integrates PINNs with both classical and quantum GANs to enable freeform design generation within a specified periodicity. The classical GANs used in this framework are conditional Deep Convolutional Generative Adversarial Networks (cDCGAN), a variant of DCGAN~\cite{Yeung2021global, radford2015unsupervised}. The QGANs used in this study are Latent Style Quantum Generative Adversarial Networks (LaSt-QGAN)~\cite{chang2024latent}.
The methodology integrates a Fano-shaped absorption spectrum equation into a Physics-Informed Neural Network (PINN) architecture, which is incorporated into both classical and quantum-enhanced Generative Adversarial Networks (GAN and QGAN, respectively). A comparative investigation is then carried out to evaluate their performance in the inverse design of metasurfaces.
Integrating spectral limitations, such as target resonance frequency, linewidth, and absorption amplitude, into the loss function directs the network towards physically meaningful solutions. The QGAN + PINN model exhibits enhanced convergence rates, superior spectrum fidelity, and requires smaller datasets relative to conventional models. This paper introduces an innovative paradigm for quantum-assisted inverse design in nanophotonics, integrating physics-informed learning with advanced quantum machine learning methodologies.

\section{Methodology}

\subsection{Framework}
\begin{figure}
    \centering
    \includegraphics[width=\linewidth]{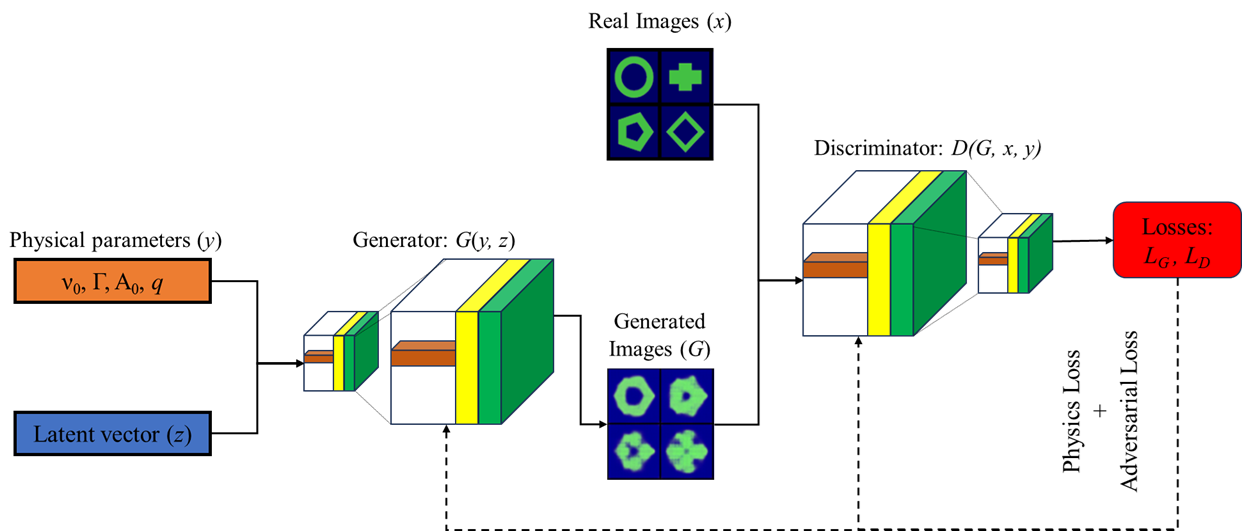}
    \caption{Schematic of the GAN + PINN architecture for inverse design of metasurface.}
    \label{fig:1}
\end{figure}

\begin{figure}
    \centering
    \includegraphics[width=\linewidth]{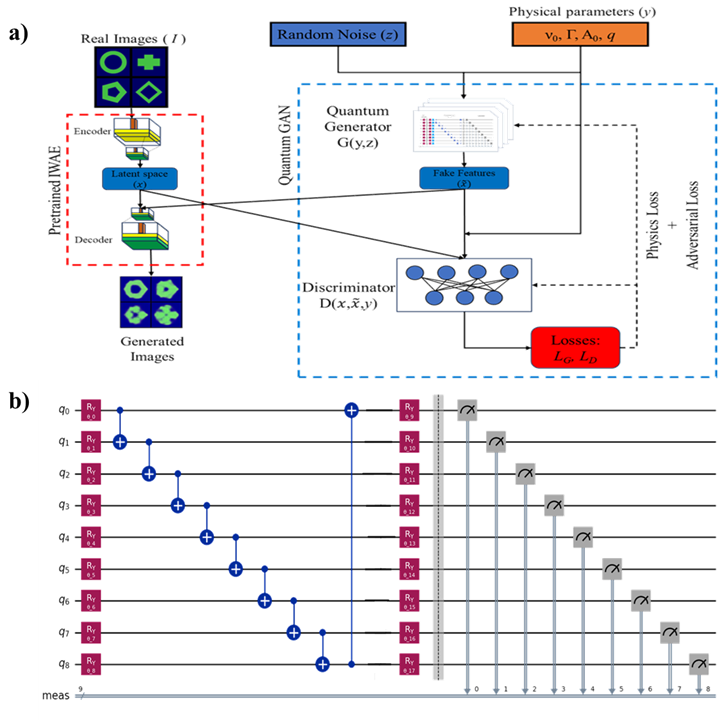}
    \caption{a) Schematic of the quantum-enhanced QGAN + PINN framework. b) Variational Quantum circuit (VQC) used in the generator for encoding latent variables.}
    \label{fig:2}
\end{figure}

To implement the inverse design of narrow-band absorbing metasurfaces, we developed a hybrid physics-informed generative adversarial learning framework.
Our approach combines a GAN, in both classical and quantum versions with a Physics-Informed Neural Network (PINN) that incorporates a Fano resonance -based absorption model, as illustrated in Figures~\ref{fig:1} and~\ref{fig:2}.
GAN is responsible for generating metasurface geometries (as 64×64 RGB images), while the PINN enforces physical constraints on the underlying optical response. The training dataset consists of:

\begin{itemize}
    \item RGB images of metasurface structures with size \(64 \times 64 \times 3\)
    \item  Physics parameters extracted from corresponding absorption spectra: resonance frequency \(\nu_0 \), linewidth \( \Gamma \), oscillator strength \(A_0\), and Fano asymmetry parameter \(q\).
\end{itemize}

The images and physics parameters are aligned such that each image represents a structure corresponding to a known set of spectral features.

The generator $G(z, \mathbf{y})$ accepts a concatenated vector of random noise $z \in \mathbb{R}^{16}$ and Fano parameters\( \mathbf{y} = [\nu_0, \Gamma, A_0, q] \), forming a latent vector. This vector is mapped through a fully connected layer and a series of transposed convolutional blocks to produce a \(64 \times 64\) RGB image. The architecture is trained to generate metasurface images whose absorption characteristics match the provided spectral parameters as shown in Figure~\ref{fig:1}.

The discriminator is a dual-headed CNN that performs the following:
\begin{itemize}
    \item Adversarial classification: (real/fake) to guide GAN training.
    \item Physics regression: predicting the 4 physics parameters from generated images.
\end{itemize}
The discriminator receives the image along with its corresponding physical parameter map and extracts features using convolutional layers before branching into two outputs.

In the quantum version (QGAN + PINN), the generator is replaced by a variational quantum circuit (VQC) as shown in the figure \ref{fig:2}a), which encodes the latent vector and physics parameters into parameterized quantum states and implemented using PennyLane simulator by Xanadu Technologies~\cite{Bergholm2018}. The physics-informed loss remains unchanged, allowing a consistent comparison between classical and quantum generative performance. The classical discriminator is a dual-headed multi-layer perceptron (MLP) with ReLU activations and a sigmoid output. A pretrained Importance-Weighted Autoencoder (IWAE) acts as a latent-space translator between images and spectra, ensuring the compatibility of generated latent states with the image decoder. A pre-trained IWAE used as a quantum generator alone is insufficient to generate \(64 \times 64\) RGB images. This limitation arises due to the inherent complexity and high dimensionality of such image data, which typically exceeds the representational capacity of standard IWAE models or current quantum hardware when used in isolation. Therefore, during the training process, input images are first encoded using a pre-trained Importance Weighted Autoencoder (IWAE), which transforms them into a compact and information-rich latent representation. This approach enables efficient learning in resource-constrained Noisy Intermediate-Scale Quantum (NISQ) devices by avoiding direct optimization over high-dimensional pixel space.

In the QGAN + PINN framework, generator is implemented as a variational quantum circuit (VQC) as shown in the figure \ref{fig:2}b), embedded within a PyTorch module using the PennyLane library. The generator accepts a 9-dimensional input vector, comprising 4 normalized physics parameters and 5 latent variables, and maps it to a feature space using quantum computation.

Each input vector $x \in \mathbb{R}^9$ is transformed using a learned orthogonal matrix $W \in \mathbb{R}^{9 \times 9}$ and a bias vector $b \in \mathbb{R}^9$, producing a parameter vector $\theta = Wx + b$. This vector serves as the rotation angles for parametrized quantum gates in the VQC.

The circuit as represented in the figure \ref{fig:2}b) consists of parameterized RY rotations applied to each qubit and cyclic entanglement via CNOT gates (qubit $i$ entangled with $i+1 \mod n$). This is followed by another layer of RY rotations repeated for a depth $d$ (i.e., number of entangling layers). After applying this ansatz, expectation values of the Pauli-Z operator for each qubit are measured as:

$$
\text{output} = \left[ \langle Z_0 \rangle, \langle Z_1 \rangle, \dots, \langle Z_{n-1} \rangle \right]
$$

These expectation values are stacked and concatenated across multiple VQC branches as defined by number of sub-generators to form the generator's output vector. This output is then decoded via a pre-trained IWAE decoder to generate the final \(64 \times 64\) RGB metasurface image. This architecture allows for efficient quantum encoding of high-dimensional design parameters and demonstrates enhanced generative capacity in low-data regimes.
The IWAE is pre-trained and kept in evaluation mode during QGAN training. VQC generator outputs are decoded via the IWAE’s image decoder. Both generator and discriminator are jointly trained using backpropagation through the hybrid architecture.

The 4 Fano parameters considered in our work - \(\nu_0 \), \( \Gamma \), \(A_0\), and \(q\) are obtained by fitting the simulated absorption spectra of each metasurface structure to Equation~\ref{eq:1} (the Fano resonance-based model). This fitting process ensures that the parameters used during training reflect physically interpretable features of the absorption line shape and directly link geometry to spectral performance. These fitted values are normalized and used both as input labels to the generator and as targets for the physics-informed loss function.

\subsection{Governing equations}

To enforce physical plausibility, we introduce a Fano resonance-based loss that evaluates the accuracy of predicted absorption spectra at the resonance frequency $\nu = \nu_0$. The absorption is modeled as:

\begin{equation}
    A(\nu) = A_0 \cdot \frac{(q + \delta)^2}{1 + \delta^2}, \quad \delta = \frac{2(\nu - \nu_0)}{\Gamma}
    \label{eq:1}
\end{equation}

Since $\nu = \nu_0 \Rightarrow \delta = 0$, this simplifies to:

$$
    A(\nu_0) = A_0 \cdot q^2
$$
This term is added to both the generator and discriminator objectives to guide learning toward physically consistent solutions.

\subsection{Physics informed Losses}

\begin{itemize}
    \item Absorption Loss ($L_A$):
    
    Penalizes the absorption ($A$) when it is less than the target absorption $A_{\text{target}}$ (e.g., 0.9).

   $$
   L_A = \left( \max\left(A_{\text{target}} - \min(A, 1.0), 0\right) \right)^2
   $$

   This is a one-sided penalty. If $A \geq A_{\text{target}}$, then there is no penalty while also making sure $A \leq 1.0$.

   \item Resonance Frequency Loss ($L_{\nu_0}$):

   Penalizes the deviation between the predicted resonance frequency $\nu_0$ and the target resonance frequency $\nu_{0,\text{target}}$.

   $$
   L_{\nu_0} = \left( \nu_0 - \nu_{0,\text{target}} \right)^2
   $$

   \item Q-factor Loss ($L_Q$):

   Penalizes the predicted Q-factor ($Q_{\text{val}}$) when it is less than the minimum target Q-factor $Q_{\min}$.

   $$
   L_Q = \left( \max(Q_{\min} - Q_{\text{val}}, 0) \right)^2
   $$

   This loss term ensures that the generated metasurface has a Q-factor greater than $Q_{\min}$. If $Q_{\text{val}} \geq Q_{\min}$, the penalty is zero.

   The quality factor $Q$ is defined as

   $$
   Q = \frac{\nu}{\Gamma} = \frac{\nu}{\Delta \nu}
   $$
\end{itemize}

The total physics-informed loss is given as:
\begin{equation}
    \mathcal{L}_{\text{physics}} = \lambda_A \cdot L_A + \lambda_{\nu_0} \cdot L_{\nu_0} + \lambda_Q \cdot L_Q
    \label{eq:2}
\end{equation}

Where, $\lambda_A$, $\lambda_{\nu_0}$, and $\lambda_Q$ are the physics hyperparameters that control the weight of each loss term and their optimzed values are given in Table~\ref{tab:1} determined by Bayesian Optimization~\cite{Mockus1989}.

\begin{table}[ht]
\caption{Optimized physics hyperparameters used in the physics loss function (Equation~\ref{eq:2}).}
\centering
\begin{tabular}{cc}
\hline
\textbf{Hyperparameter} & \textbf{Value} \\
\hline
$\lambda_A$ (Absorption loss weight) & 26 \\
$\lambda_{\nu_0}$ (Resonance frequency loss weight) & 14 \\
$\lambda_Q$ (Q-factor loss weight) & 26 \\
\hline
\end{tabular}
\label{tab:1}
\end{table}

The total loss for the discriminator is:
\begin{equation}
    \mathcal{L}_D = \mathcal{L}_{\text{D-GAN}}^{\text{real}} + \mathcal{L}_{\text{D-GAN}}^{\text{fake}} + \mathcal{L}_{\text{physics}}
\end{equation}

For the generator:
\begin{equation}
    \mathcal{L}_G = \mathcal{L}_{\text{G-GAN}} + \mathcal{L}_{\text{physics}}
\end{equation}

We use Adam optimizer with a learning rate of $10^{-5}$ and $10^{-4}$ for generator and discriminator, respectively, for batch size 16, and $\beta_1 = 0.5$. Training is carried out for 500 epochs for GAN + PINN and 250 epochs for QGAN + PINN. Real labels are sampled from a uniform distribution in $[0.9, 1.0]$, and fake labels from $[0.0, 0.1]$ to improve the stability of GAN.
The generator \(G\) and the discriminator \(D\) participate in a minimax optimization procedure.  In this context, \(G\) aims to mislead \(D\) by reducing the probability that \(D\) correctly classifies the generated data, while \(D\) attempts to increase this probability.  The objective function for this adversarial training is articulated as follows~\cite{mirza2014conditional}:

\begin{equation}
\min _G \max _D V(D, G) = \mathbb{E}_{\boldsymbol{x} \longrightarrow p_{\text{data}}}[\log D(\boldsymbol{x, y})] + \mathbb{E}_{\boldsymbol{z} \longrightarrow p_z}[\log (1 - D(G(\boldsymbol{z, y})))]
\label{eq:5}
\end{equation}
where,
\begin{itemize}
\item \(\mathbb{E}\) : Expected value
\item \(x\): Real data sample (for classical GAN it is real images \& for quantum GAN it is latent space.
\item \(z\): Random Noise
\item \(y\): Conditional vector (Radiation Profile)
\item \(D(x)\): Probability that the discriminator accurately identifies real data as authentic
\item \(G(z)\): Generated data
\item \(D(G(z))\): Probability that the discriminator categorizes generated data as authentic
\end{itemize}

\begin{figure}
    \centering
    \includegraphics[width=\linewidth]{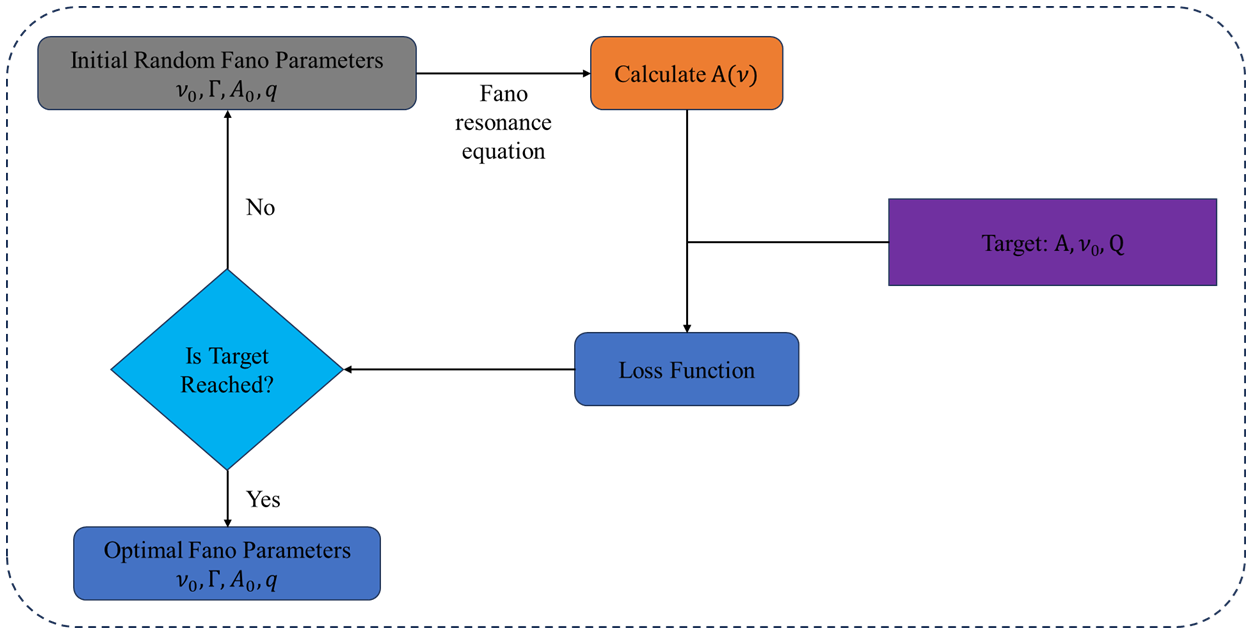}
    \caption{Flowchart of the iterative optimization process for determining optimal Fano parameters $(\nu_0, \Gamma, A_0, q)$ using a physics-informed loss function.}
    \label{fig:3}
\end{figure}

The iterative optimization is used for determining the optimal Fano parameters as shown in Figure~\ref{fig:3}. Once optimal Fano parameters are attained, the values are sent to trained classical generator for GAN + PINN, and for QGAN + PINN they are sent to trained quantum generator and pretrained decoder of IWAE to get the image of the corresponding metasurface structure.

\section{Results}

\begin{figure}
    \centering
    \includegraphics[width=\linewidth]{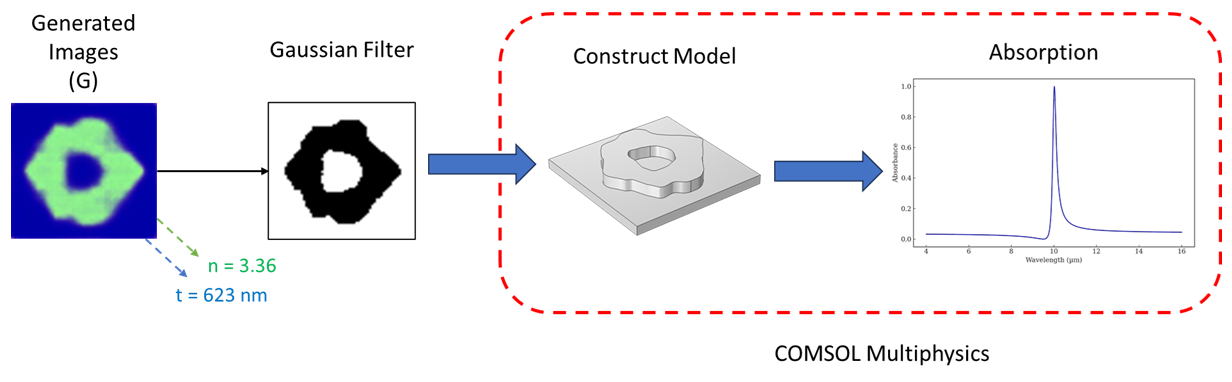}
    \caption{Workflow for computing the absorption spectrum from GAN-generated metasurface images.}
    \label{fig:4}
\end{figure}

We demonstrate the effectiveness of the proposed inverse design methodology based on PINN + QGAN by performing a comparative investigation of a sharp Fano-shaped absorption resonance using QGAN, GAN + PINN, and QGAN + PINN methodologies. The overall simulation and evaluation pipeline used in this work is illustrated in Figure~\ref{fig:4}. Firstly metasurface unit cell images are generated using the trained generative models. These images, which encode structural and spectral properties, are passed through a Gaussian filter to enhance manufacturability by smoothing irregularities. The filtered binary image is then used to reconstruct a 3D model of the metasurface in COMSOL Multiphysics. Using the physical parameters — refractive index ($n$) and thickness ($t$), predicted by the physics-informed network, the model is simulated under plane wave excitation, where the incident wave is linearly polarized along the $x$-direction to compute the absorption spectrum. This spectrum is compared against the target Fano resonance, allowing quantitative evaluation of  accuracy and physical realism of the considered design frameworks. 

\begin{figure}
    \centering
    \includegraphics[width=\linewidth]{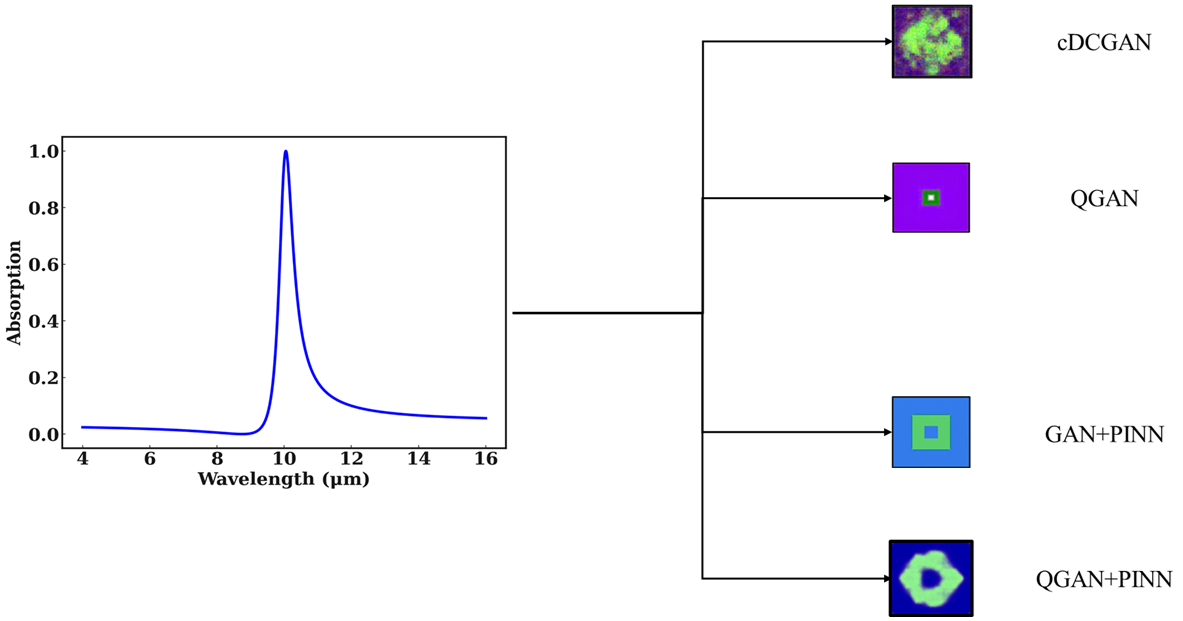}
    \caption{Inverse design outcomes from different generative models for a sharp Fano absorption spectrum centered at 10 \(\mu\)m.}
    \label{fig:5}
\end{figure}

Figure~\ref{fig:5} presents a side-by-side comparison of metasurface geometries generated by each framework in response to the same spectral target. The left panel shows the desired absorption spectrum characterized by a narrow linewidth and high asymmetry factor. This absorption profile represents a prototypical scenario requiring high spectral selectivity in mid-infrared applications such as sensing or thermal emission. All training and simulations are performed on an NVIDIA DGX-1 server equipped with 1~TB RAM, 15~TB SSD, 128 CPU cores, and a single A100 GPU, running Ubuntu 20.04. The computational efficiency demonstrated by the QGAN + PINN model on such high performance infrastructure emphasizes its scalability and potential for large scale deployment in advanced nanophotonic design tasks.

The dataset used in this study spans the mid-infrared region, covering wavelengths from 4 to 16~\(\mu\)m, corresponding to a frequency range of 18.74 to 74.95~THz. The training sample comprises 3000 metasurface unit cell patterns for the GAN + PINN system, rendered as image–vector pairings obtained from seven unique geometric designs: cross, square, ellipse, bow-tie, H, V, and tripole. In contrast, the QGAN + PINN framework is trained using a compact dataset comprising only 64 image–vector pairs, representing metasurface unit cells with plus, solid, and hollow ring geometries. These ring structures are systematically varied from 3-sided to 9-sided polygonal shapes with varying inner diameter. All dielectric metasurface patterns are embedded within square unit cells of dimensions $7.5 \times 7.5~\mu\text{m}^2$, capturing the essential structural features relevant to the absorption characteristics. Each pixel, thus represents a minimum feature size of 120~nm resolutions for dielectric, that falls well within the limits of current nanofabrication capabilities. To further improve device performance and fabrication feasibility, a Gaussian filtering post-processing step was applied~\cite{Yeung2021global}. The absorption spectra for each metasurface design, spanning the range 4–16~$\mu$m, were then obtained using finite element method (FEM) simulations carried out in COMSOL Multiphysics.

\begin{figure}
    \centering
    \includegraphics[height=0.5\linewidth,width=\linewidth]{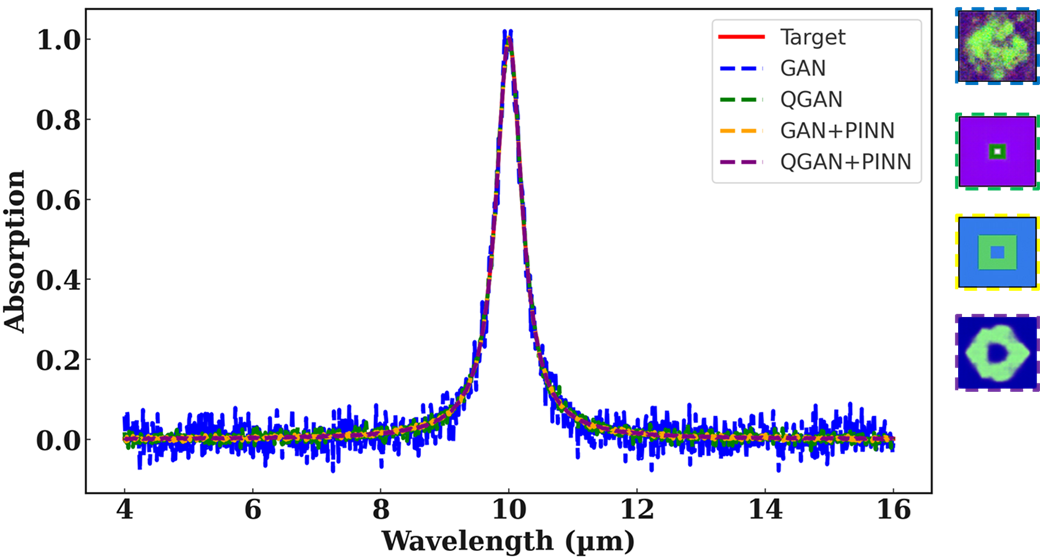}
    \caption{Comparison of absorption spectra generated by different generative models—GAN, QGAN, GAN+PINN, and QGAN+PINN—against the target Fano resonance profile (red). Each model's output spectrum is shown alongside the corresponding metasurface structure image on the right.}
    \label{fig:6}
\end{figure}

The results shown in figure \ref{fig:6} illustrate the varying performance and design quality of the four approaches.
The figure \ref{fig:6} shows absorption spectra obtained from different methods (cDCGAN, QGAN, PINN+GAN, PINN+QGAN) employed to obtain the unit cells of metasurface and the corresponding unit cells.
The cDCGAN model employed gives a blurry output for metasurface unitcell as the model is trained only on 3000 images. The QGAN-only model produces a regular geometric output even after being trained on just 500 images but lacks the refined spectral conformity due to the absence of physical constraints. In contrast, the GAN + PINN model integrates spectral physics into the learning process via the Fano-shaped loss function but requires significantly more training data (3000 samples) to generate physically plausible designs, though less than conventional GAN model, which requires atleast 20000 images to train. Finally, the QGAN + PINN architecture combines quantum-enhanced generative modeling with physics-informed learning. It produces geometries that not only conform closely to the desired spectral profile but also exhibit superior smoothness and manufacturability, while requiring far fewer training samples (only 100 images). This highlights the potential of QGAN + PINN in highly data-constrained scenarios.

The results are quantitatively summarized in Table~\ref{tab:2}, which clearly highlights the superior performance of the QGAN + PINN framework in comparison to the other models employed in this study. Specifically, QGAN + PINN demonstrates the lowest mean squared error (MSE), indicating the highest spectral accuracy in reproducing the desired absorption profile. Additionally, it achieves superior performance with the shortest training runtimes, i.e., 24 times faster than conventional GAN - as shown in Table~\ref{tab:2}. In addition, the number of training samples required to generate a clear and physically meaningful metasurface unit cell is much lower than the other methods. These combined improvements underscore the advantages of integrating quantum generative modeling with physics-informed learning, particularly in scenarios where data availability and computational resources are limited.

\begin{table}[ht]
\caption{\label{tab:2}Comparison of performance metrics across different inverse design models. *Training samples indicate the minimum number required to produce a visually clear and physically meaningful metasurface unit cell.}
\begin{tabular}{@{}lllll}
\hline
\textbf{Metric} & \textbf{GAN \cite{Yeung2021global}} & \textbf{QGAN} & \textbf{GAN + PINN} & \textbf{QGAN + PINN} \cr
\hline
Runtime (hrs)     & 25         & 2.5        & 15.08         & 1.26 \cr
MSE               & $10^{-3}$  & $10^{-4}$  & $10^{-5}$  & $10^{-6}$ \cr
Training Samples* & 20,000     & 500        & 3,000      & 100 \cr
\hline
\end{tabular}
\end{table}

To support physical realization, refractive index (\(n\)) values obtained from the inverse design are utilized to define new materials within the electromagnetic simulations for the design process of PINN or QPINN + GAN - based metasurfaces. It is important to acknowledge that traditional manufacturing methods may be insufficient for the new materials generated through this process. The framework for material definition enables a broader range of designs influenced by material properties by allowing the model to predict various material characteristics that may be overlooked due to oversimplified categories.
Unlike previous works that compute mean squared error (MSE) over the entire absorption spectrum, our approach focuses on key resonance characteristics, namely, the peak absorption, target resonance frequency, and linewidth for evaluation and comparison.

\begin{figure}
    \centering
    \includegraphics[width=\linewidth]{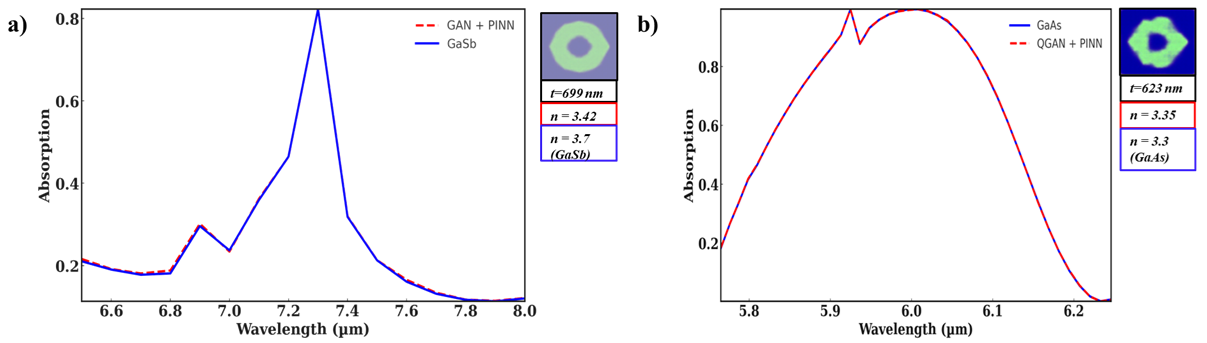}
    \caption{Comparison of absorption spectra for a) GAN + PINN and b) QGAN + PINN designed metasurfaces against reference materials (\ch{GaSb} and \ch{GaAs}, respectively). Insets show the corresponding metasurface structures and their fitted physical parameters.}
    \label{fig:7}
\end{figure}

We compare the absorption spectra of GAN + PINN and QGAN + PINN generated metasurfaces with those of reference materials (GaSb and GaAs, respectively) as demonstrated in Figure~\ref{fig:7}. The refractive index values obtained from the fitted absorption spectra are used to identify the closest-matching real materials. The inclusion of a physics-informed neural network (PINN), trained with the Fano absorption spectrum as a constraint, ensures that the generated spectra adheres to the desired resonance characteristics — target frequency or wavelength, linewidth, and peak absorption. Both models closely reproduce the target spectral response with minimal deviation with MSE of \(10^{-5}\) \& \(10^{-6}\) respectively, demonstrating high fidelity in inverse design.

\begin{figure}
    \centering
    \includegraphics[width=\linewidth]{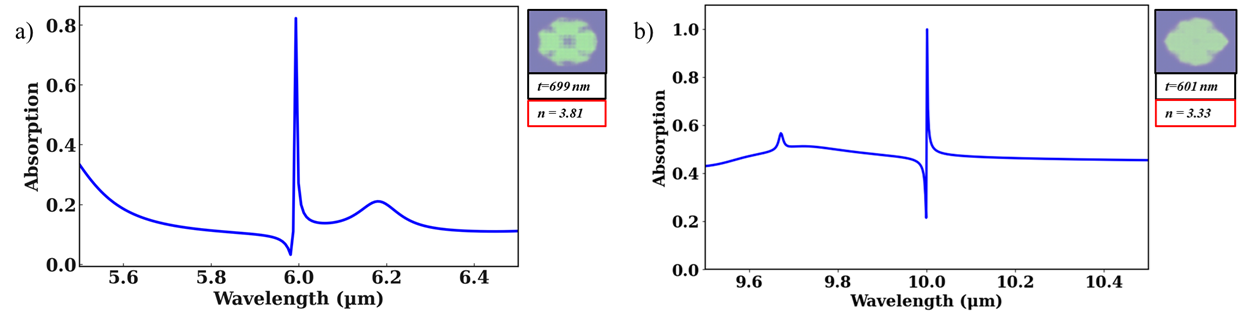}
    \caption{Absorption spectra of metasurfaces exhibiting high-Q Fano resonances at target wavelengths.}
    \label{fig:8}
\end{figure}

Further, we observe high quality factor resonances when the physics-informed loss includes the quality factor term $L_Q$, where the threshold $Q_{\min}$ is set to $10^5$. Figure~\ref{fig:8} illustrates the absorption spectra of metasurfaces generated under this constraint. In panel (a), a sharp Fano resonance with a quality factor $1.0 \times 10^5$ at 6~\(\mu\)m wavelength is observed. Panel (b) demonstrates an even sharper resonance with a Q-factor of $1.56 \times 10^5$ at 10~\(\mu\)m. The inset images show the corresponding metasurface unit cell designs responsible for producing these resonances.

Interestingly, the QGAN + PINN model tends to generate increasingly asymmetric unit cell geometries as well as ultra-high Q-factor values as the target Q-factor threshold is increased. The model implicitly learned this relationship, favoring asymmetry when the loss enforces $Q_{\min} \geq 10^5$. The dataset used for the inverse design process mostly contains Q-factor values of the order of $\sim 10^3$, which primarily correspond to symmetric metasurface structures. This observation supports the known physical intuition that highly asymmetric dielectric structures are more likely to support sharp Fano resonances with ultra high Q-factor. This emergent behavior highlights the effectiveness of combining physics-informed loss terms with quantum-assisted generative modeling for learning design rules that enhance optical quality factors.

\section{Conclusion}
In this study, we propose a novel hybrid inverse design framework that integrates physics-informed neural networks (PINNs) with both classical and quantum generative adversarial networks (GANs and QGANs) for the inverse design of narrow-band absorbing metasurfaces. While PINNs and physics-informed GANs have previously been applied to solving differential equations in mechanics \cite{PINN-GAN}, this is, to the best of our knowledge, the first application of such a framework to metasurface inverse design guided by Fano resonance equation. By embedding a Fano-shaped absorption spectrum equation directly into the loss function of the PINN, our approach ensures that the generated metasurface geometries exhibit physically accurate resonance behavior, including precise control over the target frequency, linewidth, and absorption amplitude.
Through a comparative evaluation of various performance metrics, we demonstrate that the proposed QGAN + PINN architecture significantly outperforms both the standalone GAN and the classical GAN + PINN approaches. Specifically, the QGAN + PINN model achieves lowest mean squared error (MSE), requires only 64 training samples as compared to 20000 samples for GAN, and completes training with minimal computational cost—attributes that make it especially well-suited for data-scarce or resource-constrained design scenarios. These findings (though design and application specific) highlight the unique advantages offered by quantum-enhanced generative models in efficiently navigating high-dimensional design spaces with minimal supervision.

Additionally, the integration of physics-based constraints not only improves spectral fidelity but also opens avenues for real-world applicability by allowing fitted absorption parameters to be mapped to known materials such as GaAs. This capability bridges the gap between simulation and fabrication, enabling practical deployment of designed metasurfaces in mid-infrared applications such as selective thermal emission, spectral filtering, and, notably, ultra-sensitive optical sensing.

We further observe that when the quality factor constraint $Q_{\min} \geq 10^5$ is included in the physics-informed loss, the QGAN + PINN framework consistently generates metasurfaces with sharp Fano resonances achieving Q-factors up to $1.56 \times 10^5$. These high-Q designs are accompanied by increasingly asymmetric unit cell geometries, reflecting the model’s learned preference for symmetry breaking as a mechanism to induce quasi-bound states in the continuum (quasi-BICs). Such resonances are of particular interest in sensing applications, where even small refractive index perturbations in the surrounding medium can induce measurable spectral shifts. The narrow linewidths and high sensitivity of these Fano resonances make them ideal for label-free biosensing, chemical detection, and environmental monitoring.
Although the current study focuses on narrow-band absorption and Fano resonances, the framework can be inherently extended to other functionalities. Future work may extend this platform to cover other resonance types (e.g., Lorentzian, EIT - like), material properties (e.g., dispersion, anisotropy), or structural classes (e.g., multilayer stacks, polarization-sensitive designs). This work, hence lays the foundation for a scalable, physics-grounded, and quantum-enhanced inverse design pipeline for photonic devices. The QGAN + PINN methodology not only accelerates metasurface discovery but also enables function-specific designs for applications in optics, on-chip sensing, and next-generation integrated photonic systems.

\begin{acknowledgement}
This research is supported and financed by Mahindra University.
\end{acknowledgement}

\begin{suppinfo}

\end{suppinfo}

\bibliography{achemso-demo}

\end{document}